\documentclass[aps,prl,reprint,twocolumn,superscriptaddress]{revtex4-2}
\usepackage{amssymb,amsmath,amsfonts}
\usepackage{graphicx} 
\usepackage{float} 
\usepackage{natbib} 
\usepackage{xcolor}
\usepackage[colorlinks=true,bookmarks=false,citecolor=blue,urlcolor=blue]{hyperref}
\usepackage{lmodern} 

\graphicspath{ {./Figures/} } 

\newcommand{\bk}{\mathbf{k} }

\newcommand{\bE}{\mathbf{E} }
\newcommand{\bD}{\mathbf{D} }

\newcommand{\bP}{\mathbf{P} }

\renewcommand{\Re}{\operatorname{Re} }
\renewcommand{\Im}{\operatorname{Im} }

\newcommand{\dyad}[1] {\overset\leftrightarrow{\mathbf{#1}}}

\newcommand{\mE}{ \mathcal{E} }
\newcommand{\mH}{ \mathcal{H} }

\newcommand{\eps}{\varepsilon}
\newcommand{\w}{\omega}

\newcommand{\inc}{ \mathrm{inc} }
\newcommand{\refl}{ \mathrm{refl} }
\newcommand{\tr}{ \mathrm{tr} }

\begin{document}

\title{Topological phase singularities in light reflection from non-Hermitian uniaxial media}

\newcommand{\moscow}{
Moscow Center for Advanced Studies, Moscow, 123592, Russia
}

\author{Valeria Maslova}
\affiliation{\moscow}

\author{Petr Lebedev}
\affiliation{\moscow}

\author{Denis G. Baranov}
\email[]{baranov.mipt@gmail.com}
\affiliation{\moscow}

\begin{abstract}
Perfect light transmission into a dielectric at the Brewster angle is one of the simplest effects in macroscopic electromagnetism. The common wisdom states that absorption in the dielectric violates Brewster angle and leads to a non-vanishing reflection. Yet, incorporating anisotropy may recover perfect transmission of $p$-polarized light into the absorbing medium.
Unlike the traditional "lossless" Brewster angle, perfect transmission in this case is accompanied by phase singularities of the reflection amplitude.
In this paper, we examine theoretically phase singularities and the associated topological charges emerging in the wavelength-incidence angle space upon perfect transmission into absorbing uniaxial dielectrics. We derive the analytical criterion of perfect light transmission into an anisotropic medium, demonstrate phase singularities in these scenarios, and study their dynamics as a function of material parameters.
Finally, by lowering the symmetry of the problem, we translate this phenomenon into a different parameter space of wave vector components, and illustrate the feasibility of this phenomenon with available optically anisotropic materials.
Our results may could become valuable for the development of novel analog computing schemes and holography approaches.
\end{abstract}

\maketitle
\newpage

\section{Introduction}

Perfect transmission of a $p$-polarized plane electromagnetic wave into a dielectric medium at the Brewster angle is one of the simplest yet non-trivial optical effects occurring in linear systems \cite{born2013principles}. 
Despite the simplicity, this phenomenon is of significant importance in everyday applications such as photography \cite{shen2014optical}, microscopy \cite{henon1991microscope}, and for the design of laser cavities realizing single polarization \cite{Lamb_laser}.
Brewster angle disappears in the presence of absorption in the dielectric medium. However, it was revealed recently that anisotropy of the absorbing material may recover perfect transmission of incident $p$-polarized light. A uniaxial absorbing crystal with its optical axis normal to the interface may allow perfect transmission, which is eventually accompanied by perfect absorption of the transmitted wave \cite{baranov2012electrodynamics, baranov2012perfect}. This effect has been verified using hexagonal boron nitride in the mid-IR range \cite{baranov2015perfect}.
Luo et al. have utilized this effect for the design of a broadband absorber \cite{luo2021ultra}.
Recently more exotic versions of the Brewster phenomena have been described, such as the total transmission due to a spatial homogeneity \cite{ye2017observation, horsley2015spatial}, and the temporal Brewster angle \cite{pacheco2020antireflection, pacheco2021temporal, vazquez2022shaping}. However, many these studies have overlooked the behavior of the \emph{phase} of the reflected light in the vicinity of the perfect transmission.

The amplitude of the reflected plane wave is a complex number quantified by its magnitude and argument. When the reflection magnitude drops to zero at a point of perfect absorption or transmission, the argument of the complex reflection amplitude  becomes undetermined \cite{Krasnok2019}; these points are referred to as phase singularities (PSs) \cite{hein2015retardation, ni2021multidimensional}.
Furthermore, the phase gradient may acquire a non-zero round-trip around the singularity, which gives rise to a conserved topological charge \cite{Song2017}. This non-trivial topological charge makes the phase singularity topologically protected: the winding number of the phase singularity can change only by another integer, and thus must be conserved under small variations of the scattering system.
As a side note, we emphasize that these singularities of the response function should not be confused with phase singularities of an optical field in the coordinate space, which are associated with non-zero orbital angular momentum of light \cite{Dennis2009,Dennis2010,OHolleran2009,ni2021multidimensional}.

Although previous reports mostly examined the reflection magnitude in systems featuring perfect transmission or absorption, recent studies have observed these phase singularities in  metamaterials and metasurfaces \cite{tsurimaki2018topological, yan2017phase}, periodic plasmonic nanostructures \cite{berkhout2019perfect,berkhout2020strong}, and more recently in planar systems incorporating ultra-thin films of transition metal dichalcogenides \cite{wang2020atomically, ermolaev2022topological,  canales2023perfect}.
Rapid variation of the scattered light's phase near the singularity enabled elegant applications for analog computing \cite{sol2022meta, Zangeneh-Nejad2021, Babaee2021}, image processing \cite{zhu2021topological} and molecular detection with improved sensitivity \cite{Kravets2013, vasic2014enhanced, tsurimaki2018topological, ermolaev2022topological}. 
In addition, phase singularities of higher orders has been theoretically analyzed in multi-port scattering systems \cite{liu2023spectral}.
Nonetheless, phase singularities emerging in simple uniaxial media have not been thoroughly studied yet.

In this paper, we study theoretically phase singularities emerging upon light reflection from absorbing anisotropic media. With the analytical condition of the perfect optical transmission into a uniaxial medium, we find reflection phase singularities in the parameter space of wavelength and incidence angle accompanied by non-trivial topological charges.
We examine the role of material parameters and the presence of substrate on these singularities, and find the crucial role of the material absorption of the anisotropic material for the emergence of this effect.
Finally, we identify phase singularities in the $k_x$-$k_y$ parameter space for a slanted absorbing uniaxial material with a non-vertical optical axis, and illustrate the feasibility of this phenomenon with available optically anisotropic materials.
Our results may become valuable for the development of novel optical computing devices relying on topological properties of the optical field, and holography approaches.

\section{Results}

\subsection{Total transmission in isotropic media}

We begin by briefly examining the complex reflection coefficient of a $p$-polarized plane wave incident from air ($\eps = 1$) onto an isotropic dielectric medium. In order to encompass a wider range of possible permittivities of the medium, we consider reflection from a dispersive medium described by the Lorentz model:
\begin{equation}
    \eps_\mathrm{Lor} (\w) = \eps_{\infty} + f \frac{\w_0^2}{\w_0^2 - \w^2 - i \gamma \w},
    \label{Eq_1}
\end{equation}
where $\eps_{\infty}$ is the high-frequency permittivity, $f$ is the oscillator strength  of the resonant transition of the medium, $\w_0$ is its resonant frequency, and $\gamma$ describes its non-radiative decay rate.

We first examine the case of a transparent Lorentzian dielectric with $\gamma = 0$. 
The reflection amplitude of a $p$-polarized plane wave incident at an angle $\theta$ reads:
\begin{equation}
    r_p = \frac{k_{z} - k_{z}^{(o)}/\eps(\w)}{k_{z} + k_{z}^{(o)}/\eps(\w)},
\end{equation}
where $k_{z} = \w/c \cos\theta$ and $k_{z}^{(o)} = \w/c \sqrt{\eps(\w)-\sin^2\theta}$ are the $z$-components of the wave vector of  the incident and transmitted (ordinary, $(o)$) waves, respectively.
Notice that this coefficient relates the tangential $H$-field components of two transverse-magnetic plane waves.

\begin{figure}[t!]
\centering\includegraphics[width=.5\textwidth]{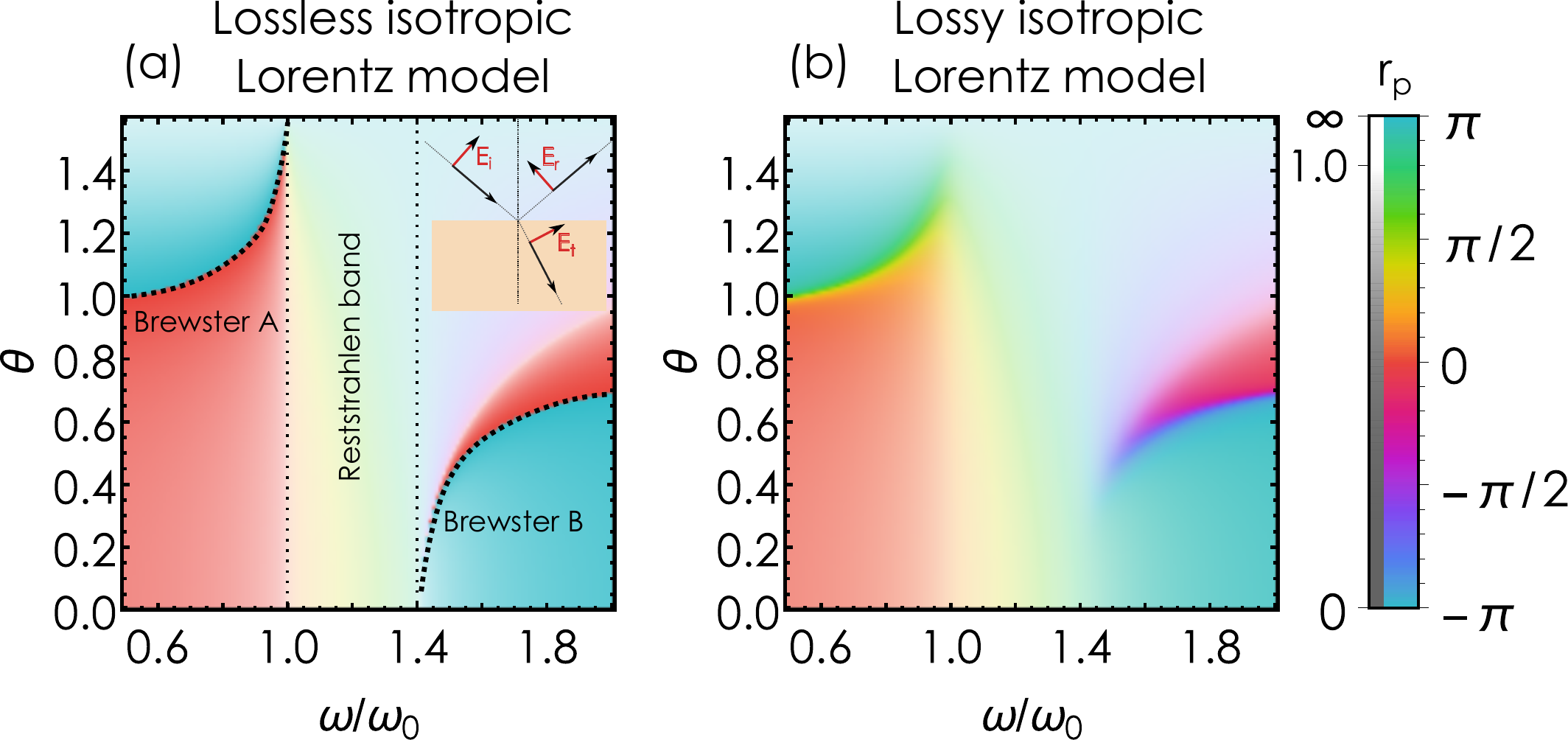}
\caption{\textbf{Perfect transmission in isotropic Lorentz media.} (a) Plot of the complex-valued reflection coefficient $r_p$ (Eq. \eqref{Eq_1}) from an isotropic Lorentz material characterized by $\eps_{\infty} = 1$, $f = 1$, $\gamma = 0$ as a function of normalized frequency and incidence angle. Hue encodes the argument of the complex amplitude. Two phase discontinuities correspond to Brewster modes A and B above and below the resonant frequency, respectively.
Inset: sketch of the scattering problem. (b) Same as (a) calculated for a non-zero material decay rate $\gamma = 0.1 \w_0$.}
\label{fig1}
\end{figure}

Figure \ref{fig1}(a) presents complex valued reflection amplitude as a function of normalized frequency and incidence angle for $\eps_{\infty} = 1$ and $f = 1$.
The plot features two discontinuity lines accompanied by a $\pi$ jump in the region of positive permittivity at $\w < \w$ and $\w > \omega_0\sqrt{1+f/\eps_{\infty}}$, Fig. \ref{fig1}(a). This discontinuity occurs exactly at the Brewster angle $\theta_B$:
\begin{equation}
    \tan \theta_B = \sqrt{\eps(\w)}.
    \label{Eq_2}
\end{equation}
Upon reflection from an isotropic Lorentz medium the reflection coefficient is real-valued below the resonant frequency of the medium, $\w < \omega_0$ at any angle, and above the $\eps = 0$ point of the medium, $\w > \omega_0\sqrt{1+f/\eps_{\infty}}$, at angles below the total internal reflection from air, $\theta < \arcsin{\sqrt{\epsilon(\w)}}$.
Furthermore, the real-valued reflection coefficient in these two regions switches its sign exactly at the Brewster angle, thus giving rise to a $\pi$ phase jump observed in the complex plot, Fig. \ref{fig1}(a).
The regions are separated by the Reststrahlen band, where permittivity is negative and there is no Brewster angle.

Adding dissipation into the Lorentzian medium, $\gamma > 0$, violates perfect transmission at the Brewster angle. Correspondingly, the phase jump disappears rendering the argument of the reflection coefficient continuous everywhere in the $\w$-$\theta$ parameter space, Fig. \ref{fig1}(b). Perfect transmission is violated because of the impedance mismatch between air and lossy medium at every incidence angle. Impedance matching can be restored by incorporating an anti-reflective coating or by placing the absorbing film on a reflective substrate, which engages the destructive interference and transforms the structure into a perfect absorber \cite{Kats2013, Kats2016, Radi2015}.

Formally Eq. \eqref{Eq_2} does admit a perfect transmission mode even with lossy media.
Indeed, plugging a complex-valued permittivity into Eq. \eqref{Eq_2} yields a complex-valued Brewster angle $\theta_B = \theta_B' + i \theta_B ''$.
This complex-valued angle describes an exponentially attenuating inhomogeneous wave.
The associated reflection coefficient zero at a complex angle can therefore be interpreted as the existence of a localized surface mode propagating at the interface of an absorbing isotropic dielectric with air \cite{frezza2015total}. Being qualitatively similar to ordinary surface-plasmon polaritons, this kind of solution is substantially weaker localized in the transverse direction and exhibits shorter propagation length, rendering it impractical for applications.

\begin{figure*}[t!]
\centering\includegraphics[width=1.\textwidth]{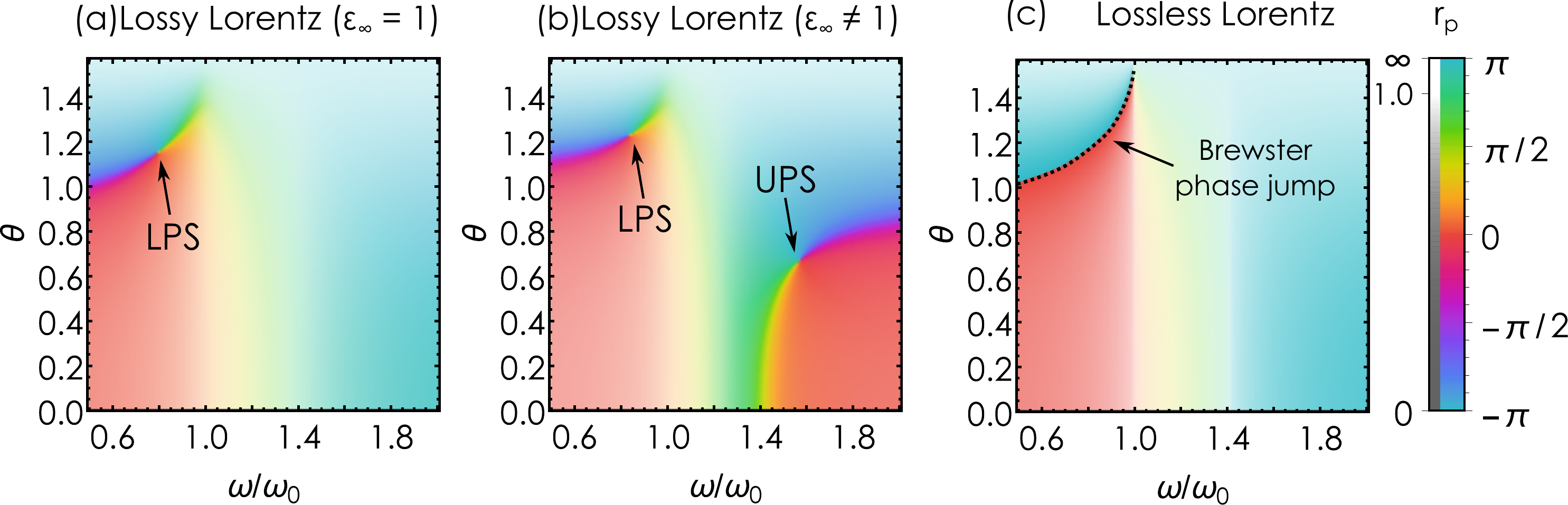}
\caption{\textbf{Perfect transmission and phase singularities in anisotropic Lorentz media.} (a) Plot of the complex-valued reflection coefficient $r_p$ (Eq. \eqref{Eq_1}) from a uniaxial material characterized by $\eps_{\infty} = 1$, $f = 1$, $\gamma = 0.1 \omega_0$, and $\eps_{\parallel} = 2 + 0.5i$ as a function of normalized frequency and incidence angle. The (lower) phase singularity (LPS) occurs at the point of perfect transmission of p-polarized wave into the absorbing uniaxial medium.
(b) Same as (a) plotted for $\eps_{\infty} = 2$. Another point of perfect transmission occurs above the Lorentz resonance accompanied by the upper phase singularity (UPS). 
(c) Same as (a) calculated for a lossless uniaxial medium with $\gamma = 0$ and $\eps_{\parallel} = 2$. No phase singularities are observed, and the Brewster mode accompanied by a phase discontinuity is restored.}
\label{fig2}
\end{figure*}

\subsection{Total transmission in anisotropic media}

Alternatively to engaging the destructive interference, total transmission of a homogeneous wave illuminating an absorbing medium can be restored by making that medium anisotropic. It was shown theoretically that certain absorbing plasmonic metamaterials with their optical axis perpendicular to the interface allow perfect transmission of a $p$-polarized incident wave \cite{baranov2012electrodynamics, baranov2012perfect}.
In the following we are going to examine a more general class of absorbing uniaxial media. 
Consider a uniaxial material with the following permittivity tensor
\begin{equation}
\dyad{\eps} = 
    \begin{pmatrix}
        \eps_{xx} & 0 & 0 \\
        0 & \eps_{yy} & 0 \\
        0 & 0 & \eps_{zz}
    \end{pmatrix},
    \label{Eq_eps_tensor}
\end{equation}
where $\eps_{xx} = \eps_{yy} \equiv \eps_{\bot}$ is described by the Lorentz model, Eq. \eqref{Eq_1}, while the permittivity along the optical axis is a complex-valued constant, $\eps_{zz} \equiv \eps_{\parallel} = \mathrm{const}$. 
This choice of the permittivity tensor model is motivated by the abundance of naturally occurring highly anisotropic van der Waals materials, many of which feature resonant Lorentzian response along the in-plane direction, and a non-resonant or a weakly dispersive permittivity along the optical axis \cite{li2014measurement, Ma2018, Hu2017, Ermolaev2020, Ermolaev2021, ling2021all, munkhbat2022optical}.

Reflection coefficient of a $p$-polarized plane wave illuminating a uniaxial medium with its optical axis perpendicular to the interface takes the form:
\begin{equation}
    r_p = \frac{ k_{z} - k_{z}^{(e)}/\eps_\perp(\w) }{ k_{z} + k_{z}^{(e)}/\eps_\perp(\w) },
\label{Eq_3}
\end{equation}
where $k_{z} = \omega/c \cos\theta$, and $k_{z}^{(e)} = \omega/c \sqrt{\eps_\perp(\omega)/\eps_\parallel (\eps_\parallel - \sin^2\theta)}$ is the $z$-component of the wave vector of the $p$-polarized (extraordinary, $(e)$) transmitted wave in the now uniaxial medium.
Figure \ref{fig2}(a) shows the resulting complex reflection amplitude of a $p$-polarized wave incident on a uniaxial material with some generic parameters ($\eps_\infty = 1$, $f = 1$, $\gamma = 0.1 \omega_0$,  and $\eps_{\parallel} = 2 + 0.5i$). The most striking observation of this calculation is the point of zero reflection just below the resonant frequency accompanied by a phase singularity (PS). The PS can be quantified by the winding number:
\begin{equation}
    C = \frac{1}{2\pi} \oint \nabla (\arg (r_p)) \cdot d\mathbf{l},
    \label{Eq_4}
\end{equation}
where the integration contour encloses a phase singularity.
For the particular case of an absorbing Lorentz uniaxial medium with $\eps_\infty = 1$ the winding number of the reflection PS appears to be equal to $+1$  for any choice of other material parameters.

Remarkably, replacing the high-frequency dielectric constant $\eps_\infty$ by a value different from 1 adds another PS on the other side of the resonance, $\omega > \omega_0$, with an opposite topological charge, Fig. \ref{fig2}(b). We refer to the two resulting phase singularities as the lower phase singularity (LPS) and upper phase singularity (UPS), respectively.

Now we go back to the case of $\eps_\infty = 1$ and analyze the effect of dissipation on the presence of PSs in the $\omega$-$\theta$ parameter space.
Interestingly, removing the dissipation form the uniaxial material (by setting $\gamma = 0$ and $\eps_\parallel = 2$) preserves total transmission from air accompanied by the phase jump, but removes the PS entirely, Fig. \ref{fig2}(c).
In other words, the presence of non-zero dissipation in the anisotropic material is required for non-trivial topological phase singularities in reflection.

\begin{figure}[t!]
\centering\includegraphics[width=.5\textwidth]{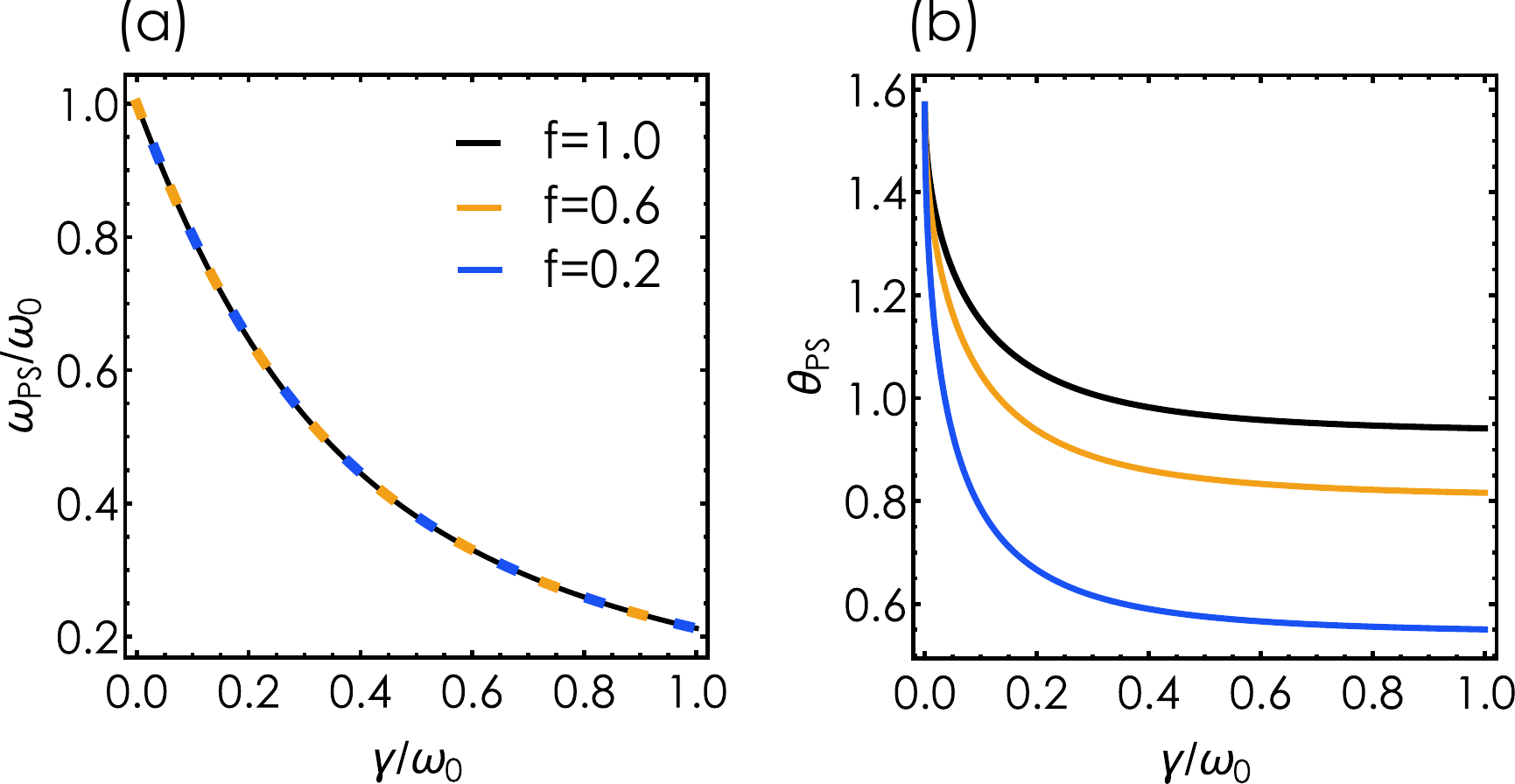}
\caption{\textbf{The effect of material parameters on phase singularities in media with $\eps_{\infty}=1$.} 
(a) Frequency of the phase singularity as a function of the resonance linewidth $\gamma$ for a series of oscillator strengths. The permittivity along the optical axis $\eps_{\parallel} = 2+0.5i$. (b) Same as (a) for the corresponding angle of the phase singularity.
}
\label{fig3}
\end{figure}

\begin{figure*}[t!]
\centering\includegraphics[width=0.67\textwidth]{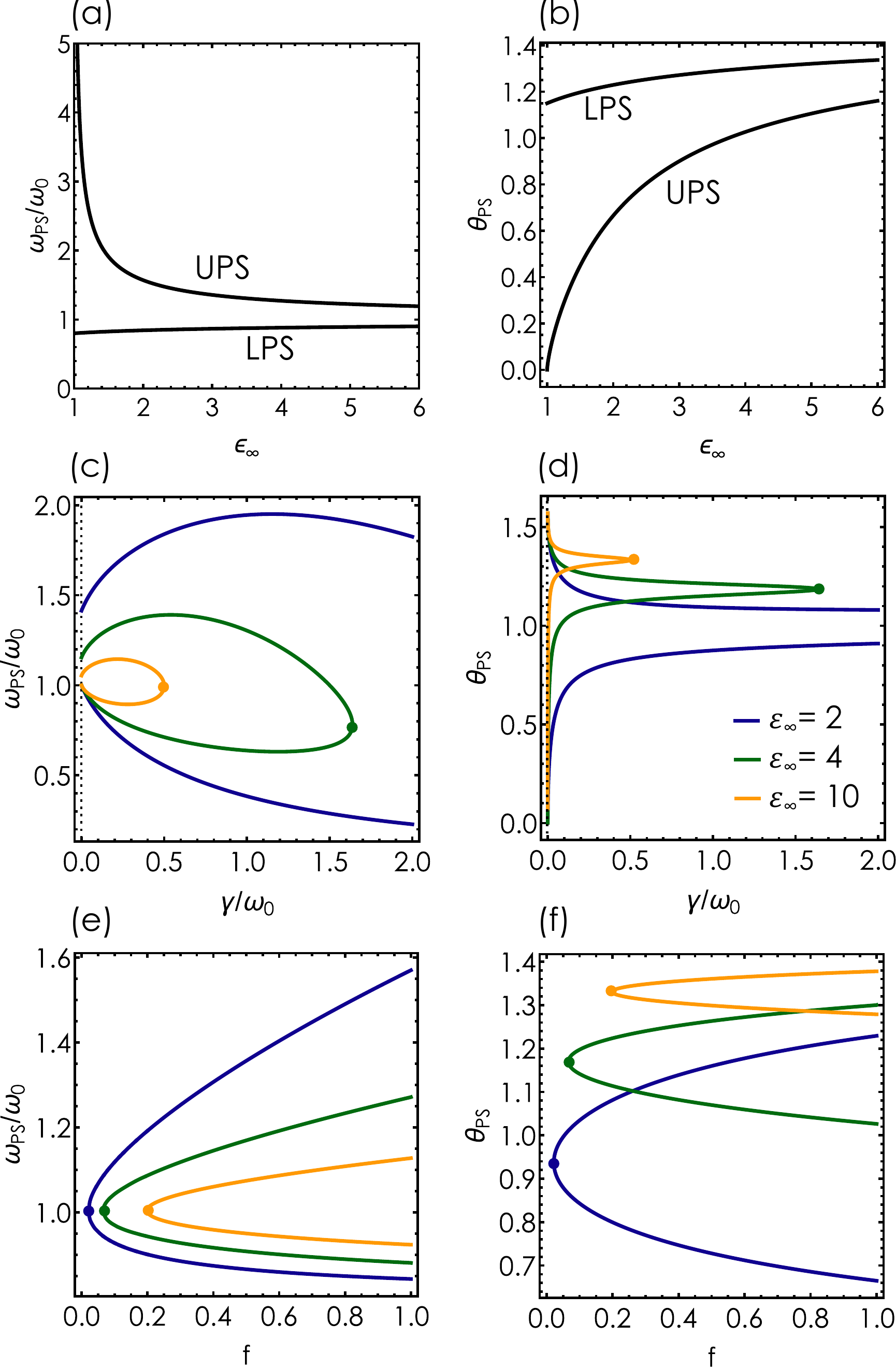}
\caption{\textbf{The effect of material parameters on phase singularities in media with $\mathbf{\eps_{\infty}\neq 1}$.} 
\footnotesize {(a) Frequency and (b) angle of the two phase singularities as a function of the high-frequency dielectric constant $\eps_\infty$ for $f=1$, $\gamma = 0.1\omega_0$, and $\eps_\parallel = 2+0.5i$.
(c) Frequency and (d) angle of the two reflection phase singularities as a function of the Lorentz resonance linewidth for different values of $\eps_\infty$. Dots indicate the points of annihilation of two oppositely charged upper and lower phase singularities.
(e) Frequency and (f) angle of the two reflection phase singularities as a function of the Lorentz oscillator strength for different values of $\eps_\infty$.}
}
\label{fig4}
\end{figure*}

Equation \eqref{Eq_3} allows us obtaining an analytical condition of total transmission upon reflection by a uniaxial material. Equating the numerator of Eq. \eqref{Eq_3} to zero yields:
\begin{equation}
    \cos^2 \theta = \frac{\eps_\parallel - 1}{\eps_\parallel \eps_{\perp} - 1}.
    \label{Eq_5}
\end{equation}
The latter yields a real-valued incidence angle if and only if
$\Im [\cos^2{\theta}] = 0$ and
$0 < \Re [\cos^2{\theta}] < 1$.
Plugging the dispersive components $\eps_{\bot}$ and $\eps_{\parallel}$ of our permittivity tensor into Eq.~\ref{Eq_5} and splitting into real and imaginary parts, from the first condition we arrive at the equation for the frequency of zero reflection points:
\begin{equation}
\begin{split}
    \eps_\parallel ''(\eps_{\infty}-1) \left( \frac{\w_\mathrm{PS}}{\w_0} \right)^4 + \\
    \eps_\parallel'' \left( (\eps_{\infty} - 1) \left( \left( \frac{\gamma}{\w_0} \right)^2 - 2 \right) - f\right) \left(\frac{\w_\mathrm{PS}}{\w_0}\right)^2 - \\
    f \frac{\gamma}{\w_0} \left( |\eps_\parallel|^{2} - \eps_\parallel ' \right) \left(\frac{\w_\mathrm{PS}}{\omega_0}\right) + \eps_\parallel ''(\eps_{\infty}+f-1) 
    = 0.
\end{split}
    \label{Eq_8}
\end{equation} 
In the case of $\eps_{\infty}=1$ this equation is simplified and the frequency is determined by 
\begin{equation}
\begin{split}
    \frac{2 \omega_\mathrm{PS}}{\w_0} =      
    -\left(\frac{\gamma}{\w_0} \right)\frac{|\eps_\parallel|^{2} - \eps_\parallel'}{\eps_\parallel''} + \\
    \sqrt{ \left(\frac{\gamma}{\w_0} \right)^2 \left( \frac{|\eps_\parallel|^{2} - 
    \eps_\parallel'}{\eps_\parallel''} \right)^{2} + 4 }.
\end{split}
    \label{Eq_9}
\end{equation}
Finally, plugging the latter into Eq.~\eqref{Eq_5} one obtains the corresponding incidence angle of perfect transmission.

Figure \ref{fig3} shows the PS frequency and incidence angle as a function of the decay constant $\gamma$ for a series of oscillator strength $f$ for the extreme case $\eps_\infty = 1$.
In this case the problem features a single PS, whose frequency does not depend on the oscillator strength, but does depend on the resonance linewidth, \ref{fig3}(a).
The corresponding incidence angle is different for various $f$, and approaches $\pi/2$ (grazing incidence) in the limit of vanishing material decay rate $\gamma \to 0$.

Next we analyze these behaviors in the case of $\eps_{\infty} \ne 1$. Although the analytical expressions in this case are quite cumbersome, dependencies plotted in Fig.~\ref{fig4} offer  the overall understanding of the PSs behavior.
In the case of a high-frequency permittivity other than unity, two perfect transmission points accompanied by PSs with opposite unitary topological charges coexist. 
As $\eps_\infty$ approaches 1, the LPS red-shifts slightly, while the UPS blue-shifts and eventually disappears at $\eps_\infty = 1$, Fig.~\ref{fig4}(a). Both PSs acquire larger incidence angles with increasing $\eps_\infty$, Fig.~\ref{fig4}(b).
As $\gamma$ approaches zero for a finite dissipation along the optical axis, the angles of the lower and upper PSs approach $\pi/2$ and 0, respectively, Fig.~\ref{fig4}(d).
Notably, there are material parameters at which the two PSs merge into one and annihilate, as marked by circles in Figs.~\ref{fig4}(c)-(f). 
The position of annihilation points is sensitive to the material parameters, and the process of merging of the UPS and LPS can be observed at fairly realistic parameters (see Fig. S1 of Supporting Information).

The phenomenon of Brewster angle in \emph{isotropic} dielectrics is often interpreted at a microscopic level by considering the polarization density $\bP$ of the transmitted wave in the medium. 
In the described scenario of a $p$-polarized plane wave incident from air at an angle $\theta$ the transmitted wave propagates at an angle $\theta_t$ related to the incidence angle by the Snell's law: $\sin \theta / \sin \theta_t = \sqrt{\eps}$. The polarization of the transmitted wave, at the same time, is perpendicular to the wave vector and makes an angle $\pi/2 - \theta_t$ with the normal. 
Loosely speaking, the reflected wave is produced by the oscillations of the polarization density -- the collection of induced point dipoles -- in the second medium. For a plane wave excitation at an angle $\theta = \arctan \sqrt{\eps}$ the propagation direction of the reflected wave happens to match exactly the direction of the axis of the induced dipoles in the second medium. However, a point dipole does not radiate along its axis, thus completely prohibiting any radiation in the direction of the reflected wave.

We confirm that the phenomenon of total transmission into an absorbing anisotropic material allows the same interpretation. To that end, we calculate the degree of orthogonality $\delta$ between the macroscopic polarization and the wave vector of the reflected wave:
\begin{equation}
    \delta =  \left| \frac{\bP \times \bk_{\refl}}{|\bP| \cdot |\bk_{\refl}| } \right|,
\end{equation}
where $\bk_{\refl} = (k_x, 0, -k_z)$ is the wave vector of the reflected wave.
Macroscopic polarization itself is not perpendicular to the wave vector inside the anisotropic medium. However, we can easily relate it to the displacement via
\begin{equation}
    \bP = \eps_0 (\dyad{\eps} - 1) \bE = (\dyad{\eps} - 1) \dyad{\eps}^{-1} \bD,
\end{equation}
where $\bD$ lies in the $xz$ and is perpendicular to $\bk_{\tr} = (k_z, 0, k_z^{(e)})$ inside any medium thanks to $\nabla \cdot \bD = 0$. 
Thus, by inverting the permittivity tensor of the uniaxial medium we easily relate the macroscopic polarization to the wave vector of the transmitted wave.

\begin{figure*}[t!]
\centering\includegraphics[width=1\textwidth]{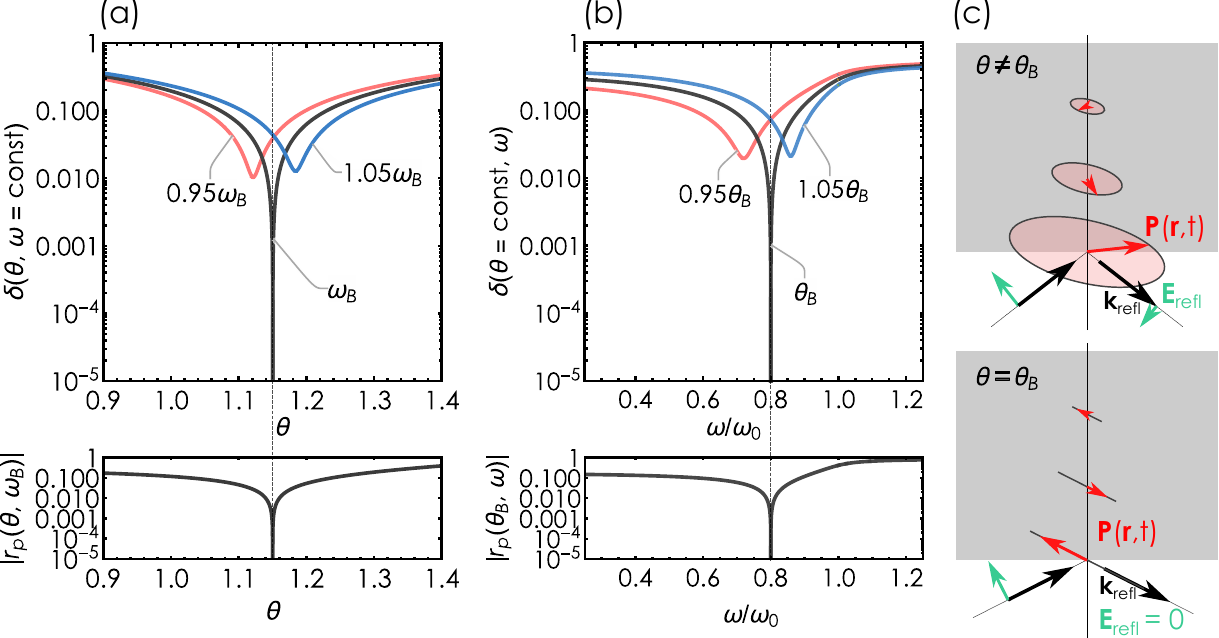}
\caption{\textbf{Microscopic picture of perfect transmission at the Brewster angle.} 
(a) The degree of orthogonality $\delta$ between the polarization density of the transmitted $p$-polarized wave and the wave vector of the reflected wave as a function of the incident angle at a fixed wavelength of phase singularity. The data is calculated for  $\eps_{\infty} = 1$, $f = 1$, $\gamma = 0.1 \omega_0$, and $\eps_{\parallel} = 2 + 0.5i$. Bottom: the corresponding reflection amplitude $|r_p|$. 
(b) Same as (a) but plotted as a function of frequency of the incident $p$-polarized wave at the angle of phase singularity. 
(c) Schematic illustration of the microscopic mechanism of total transmission in absorbing uniaxial media. At an angle different from the angle of perfect transmission $\theta_{PS}$ the incident field induces a transmitted wave with a complex-valued vector of polarization $\bP$ describing an elliptical polarization, which in turn produces a reflected wave. At an angle of perfect transmission $\theta_{PS}$, the induced polarization becomes linear and parallel with the wave vector of the reflected wave, thus forbidding any reflected field.}
\label{fig5}
\end{figure*}

As expected, this degree of orthogonality (calculated either as a function of incident angle at a fixed wavelength of phase singularity, Fig. \ref{fig5}(a), or vice versa, Fig. \ref{fig5}(b)) drops to zero exactly at the point of perfect transmission. This confirms that perfect transmission of a p-polarized wave into an absorbing uniaxial medium is accompanied by alignment of the medium polarization with the wave vector of the reflected wave, as sketched schematically in Fig. \ref{fig5}(c).
In Supplementary Note S1 we provide a general proof of equivalence between zero reflection of a p-polarized wave and vanishing degree of orthogonality $\delta$ for this particular arrangement with vertical optical axis.



To conclude this subsection, we examine the presence of topological PSs in the inverted case with the Lorentz model describing the permittivity of the uniaxial material along its optical axis: $\eps_\parallel (\omega) = \eps_{\mathrm{Lor}}$, and $\eps_\perp = \mathrm{const}$. 
Although natural materials with a resonant response along the optical axis in the visible range are far less common (if not nonexistent), artificial metamaterials allow obtaining highly anisotropic effective permittivity tensors, in particular with resonant behavior along the optical axis \cite{smith2003electromagnetic, smith2004negative, Poddubny2013, shekhar2014hyperbolic, ferrari2015hyperbolic}.

\begin{figure*}[t!]
\centering\includegraphics[width=1.\textwidth]{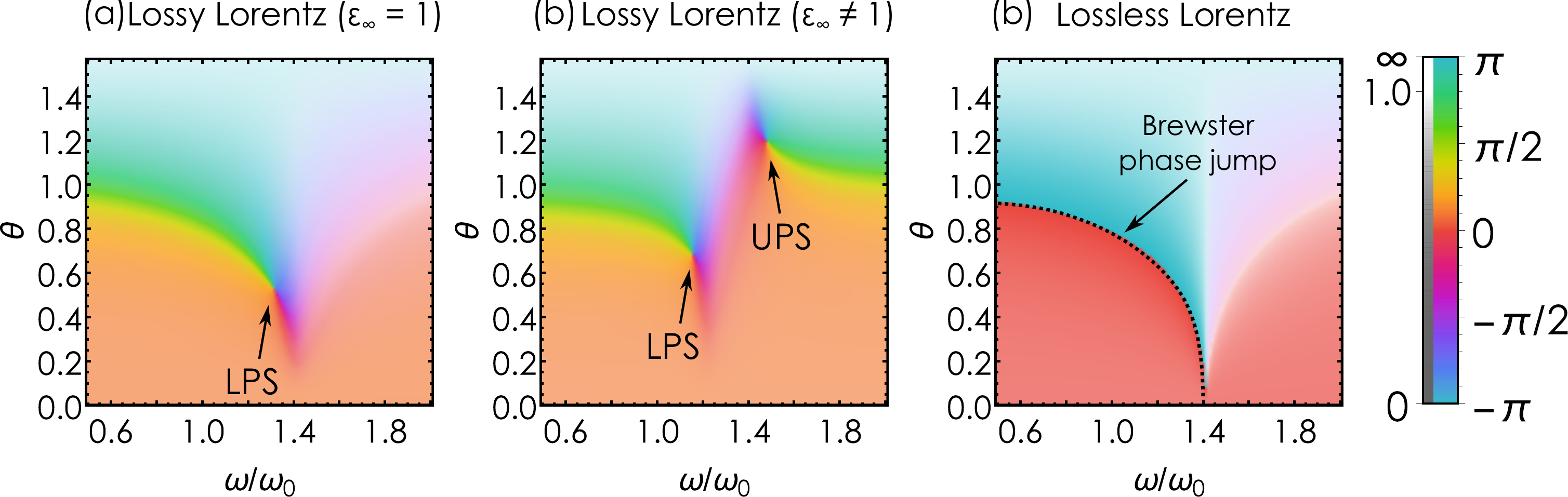}
\caption{\textbf{Phase singularities in uniaxial media with Lorentz response along the optical axis.} (a) Complex-valued reflection coefficient $r_p$ from a uniaxial material with $\eps_{\infty} = 1$, $f = 1$, $\gamma = 0.1 \omega_0$, and $\eps_{\perp} = 2 + 0.5i$ as a function of normalized frequency and incidence angle. (b) Same as (a) for $\eps_{\infty} = 2$. (c) Same as (a) calculated for a lossless uniaxial medium with $\gamma = 0$ and $\eps_{\parallel} = 2$. The phase discontinuity is observed at the Brewster angle.}
\label{fig6}
\end{figure*}

Plot of the complex-valued reflection amplitude of a $p$-polarized plane wave calculated for $f = 1, \gamma = 0.1$, $\eps_\infty = 1$, $\eps_{\perp} = 2 + 0.5i$ confirms the presence of a topologically protected PS in this scenario, Figure \ref{fig6}(a).
Similarly to the case of the in-plane Lorentzian response, in the case of $\eps_{\infty} \neq 1$ the problem features two phase singularities with opposite topological charges corresponding to the LPS and UPS and only LPS remains otherwise. 
The topological charges of LPS and UPS have opposite signs compared to the case considered above. Removing dissipation from the system preserves total transmission accompanied by the phase jump at the Brewster angle but violates the phase singularity as previously. 

The inverted model also allows us obtaining the equation for the PS frequency and angle analogous to Eqs. \eqref{Eq_8} and \eqref{Eq_9}, and analyzing the behavior of the PS as a function of material parameters. 
Fig. S3 of Supporting Information shows an example of these dependencies versus the screening constant $\eps_\infty$, the resonance decay rate $\gamma$, and the oscillator strength $f$. Overall, the resulting dependencies are somewhat similar to the case of the in-plane Lorentz response, but with quantitative differences owing to the different geometry of the problem.

\subsection{Finite thickness film}

Although semi-infinite anisotropic medium is a useful analytical model that allows us to establish the basics of PSs behavior in uniaxial absorbing media, all such materials typically exist in the form of relatively thin films deposited on a dielectric substrate.
For this reason we now briefly examine the appearance of topological PSs in finite thickness films of uniaxial absorbing media on top a dielectric substrate.

In the case of a $p$-polarized incident wave and $\eps_\infty = 1$, for wavelength-thick films the problem features one PS  that originates from the dissipative Brewster condition, Eq. \eqref{Eq_5}, and whose position in the $\omega$-$\theta$  space is shifted compared to the case of a semi-infinite medium (see Fig. S4(b) of Supporting Information). 
For thicker films new singularities associated with the Fabry-Perot resonance of the film appear at frequencies above $\omega_0$ (see Fig. S4(c) of Supporting Information). 
The situation is different for the high-frequency permittivity different from unity ($\eps_\infty = 2$), where an additional phase singularity appears already for a wavelength-thick film.
At the same time, no additional PSs appear for thicker films (see Fig. S4(f) of Supporting Information).

In the case of an $s$-polarized incident wave, finite film thickness allows observing PSs in contrast to the problem of a semi-infinite medium.
Their behavior overall repeats the behavior of additional PSs in the case of a $p$-polarized illumination (see Fig. S5 of Supporting Information).

\subsection{Slanted anisotropic media: $k_x$-$k_y$ parameter space}

\begin{figure}[b!]
\centering\includegraphics[width=.5\textwidth]{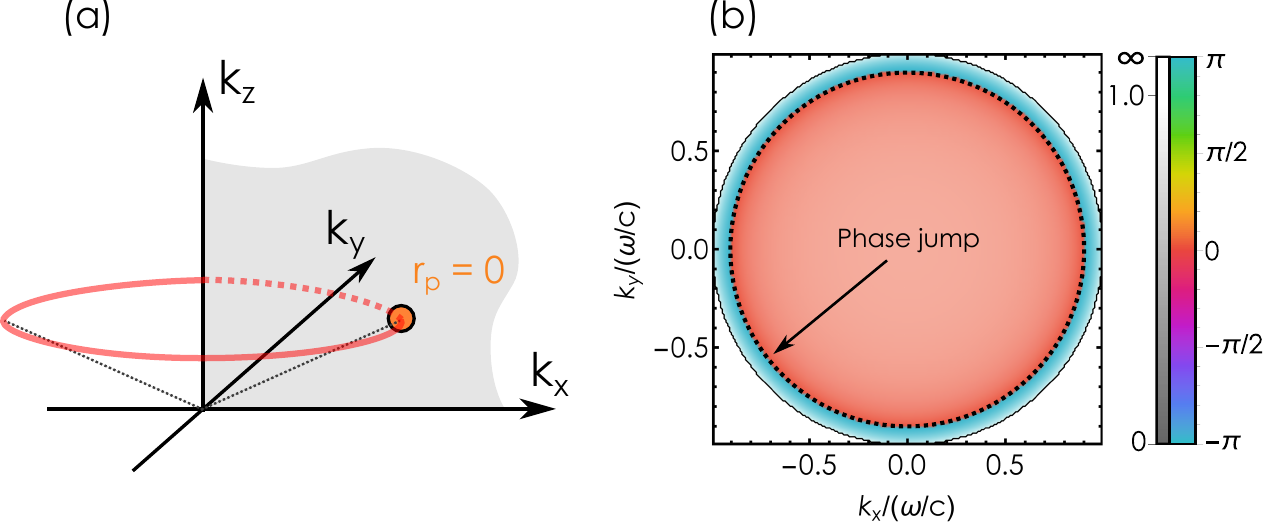}
\caption{\textbf{Phase singularities visualized in various parameter spaces.} (a) The loop of phase singularities embedded in the full $\bk$-space. Intersection of the loop with the $k_x$-$k_z$ space (equivalent to the $\w$-$\theta$ space) yields a single point of zero reflection accompanied by a phase singularity.
(b) Complex-valued reflection coefficient $r_p$ (Eq. \eqref{Eq_3}) as a function of $k_x$ and $k_y$ evaluated at a fixed frequency $\w = \w_\mathrm{PS}$ for $f = 1, \gamma = 0.1$, $\eps_\infty = 1$, $\eps_\parallel = 2 + 0.5i$. The argument features a circular discontinuity at the values corresponding to $\theta = \theta_{PS}$, and no isolated phase singularities.}
\label{fig7}
\end{figure}

So far we have examined reflection PSs in $\omega$-$\theta$ parameter space of wavelength (frequency) and incidence angle for a fixed vertical incidence plane, which is equivalent to a single quadrant of the $k_x$-$k_z$ space:
\begin{equation}
    k_x = \frac{\omega}{c} \sin \theta, \quad
    k_z = \frac{\omega}{c} \cos \theta.
\end{equation}
This parameter space, in turn, is a subspace of the bigger parameter space described by the Cartesian components $k_x$, $k_y$, and $k_z$ of the wave vector of the incident wave.
Thanks to the axial symmetry of the configuration examined above, the observed PS is a part of an entire \emph{loop} of singularities embedded in the full $\bk$-space, Fig.~\ref{fig7}(a).
Projecting this loop of singularities onto the $\omega$-$\theta$ space yields exactly a single point that we observed in the calculations presented in Fig.~\ref{fig2}(a).

\begin{figure*}[t!]
\centering\includegraphics[width=1.0\textwidth]{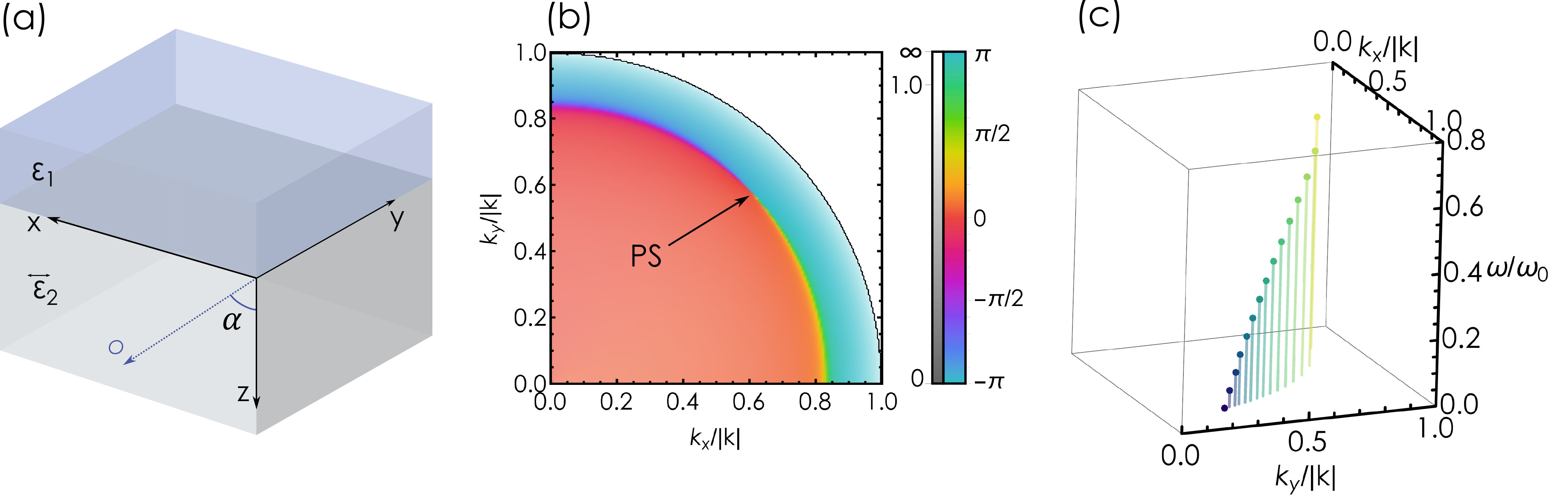}
\caption{\textbf{Phase singularities of slanted uniaxial media.} 
(a) Sketch of the scattering problem. (b)  Complex-valued reflection coefficient $r_{pp}$ from a uniaxial material characterized by $\sin{\alpha}=0.6$,  $\eps_{\infty} = 1$, $f = 1$, $\gamma = 0.1 \omega_0$, and $\eps_{\parallel} = 2 + 0.5i$ as a function of normalized wave vectors $k_x/|\bk|$ and $k_y/|\bk|$ for a fixed frequency $\omega=0.5\omega_{0}$. (c) Trajectory of phase singularity in the $\bk_\parallel$-$\omega$  space. I also added the figure in Supplementary Note 4 about the case of Lorentz response along the optical axis.
}
\label{fig8}
\end{figure*}

In the context of practical applications such as analog computing and image processing \cite{zhu2021topological}, it would be advantageous to have at our disposal a system exhibiting an isolated PS in the $k_x$-$k_y$ space at a fixed frequency (wavelength).
This would allow us to process an incident monochromatic field composed of various wave vectors depending on their propagation direction (spatial frequency).
The available $k_x$-$k_y$ space at a fixed frequency describes a 2-disk $k_x^2 + k_y^2 < (\omega/c)^2$.
Correspondingly, plotting the argument of the reflected $p$-polarized wave in this parameter space images the entire singularity loop with a phase jump across the loop, Fig.~\ref{fig7}(b).
This poses a quest for structures exhibiting isolated phase singularities in the $k_x$-$k_y$ space.

For such a point to become possible, the axial symmetry of the system with respect to the vertical axis must be broken. This can be accomplished by considering a biaxial material with $\eps_{xx} \ne \eps_{yy}$. Resonant biaxial media exist in nature and exhibit resonant permittivity components \cite{munkhbat2022optical,shubnic2020high,singh2020refractive}. However, they are relatively rare and come with a vast space of parameters controlling their permittivity components.
On the other hand, we can violate the axial symmetry with a uniaxial material by rotating it by an angle $\alpha$ around the horizontal axis. 
Without loss of generality we perform the rotation around the y-axis as shown in Fig.~\ref{fig8}(a). The permittivity of the rotated uniaxial medium takes the form:
\begin{equation}
    \dyad{\eps}' = \dyad U \dyad{\eps} \dyad U^{-1},
    \label{eps'}
\end{equation}
where $\dyad U$ is the rotation matrix around the y-axis:
\begin{equation}
    \dyad U =
    \begin{pmatrix}
        \cos{\alpha} & 0 & \sin{\alpha} \\
        0 & 1 & 0 \\
        -\sin{\alpha} & 0 & \cos{\alpha}
    \end{pmatrix}.
\end{equation}

The interaction of incident plane wave with the slanted uniaxial crystal does no longer preserve polarization states of $s$- and $p$- polarized light.
The \emph{scalar} amplitudes of the reflected s- and p- waves are related to the ones of the incident field via (see Note S4 of Supplemental Material for the details of the derivation):
\begin{equation}
    \begin{pmatrix}
        \mE_{\refl}^{(s)} \\
        Z_0 \mH_{\refl}^{(p)}
    \end{pmatrix} = \dyad{R}
    \begin{pmatrix}
        \mE_{\inc}^{(s)} \\
        Z_0 \mH_{\inc}^{(p)}
    \end{pmatrix},
\end{equation}
where $\dyad R$ is the \emph{reflection matrix}:
\begin{equation}
    \dyad R = \begin{pmatrix}
    r_{ss} & r_{sp}\\
    r_{sp} & r_{pp}
    \end{pmatrix}.
\end{equation}
Here, the diagonal elements $r_{ss}$ and $r_{pp}$ characterize reflection maintaining polarization, while the off-diagonal elements $r_{sp}$ and $r_{ps}$ describe reflection with polarization conversion.
We will focus of the $r_{pp}$ amplitude describing reflection of a $p$-polarized wave back into $p$-polarized one, since it continuously transforms into $r_p$ reflection coefficient with $\alpha \to 0$, which is the subject of our previous analysis.

\begin{figure*}[ht!]
\centering\includegraphics[width=0.9\textwidth]{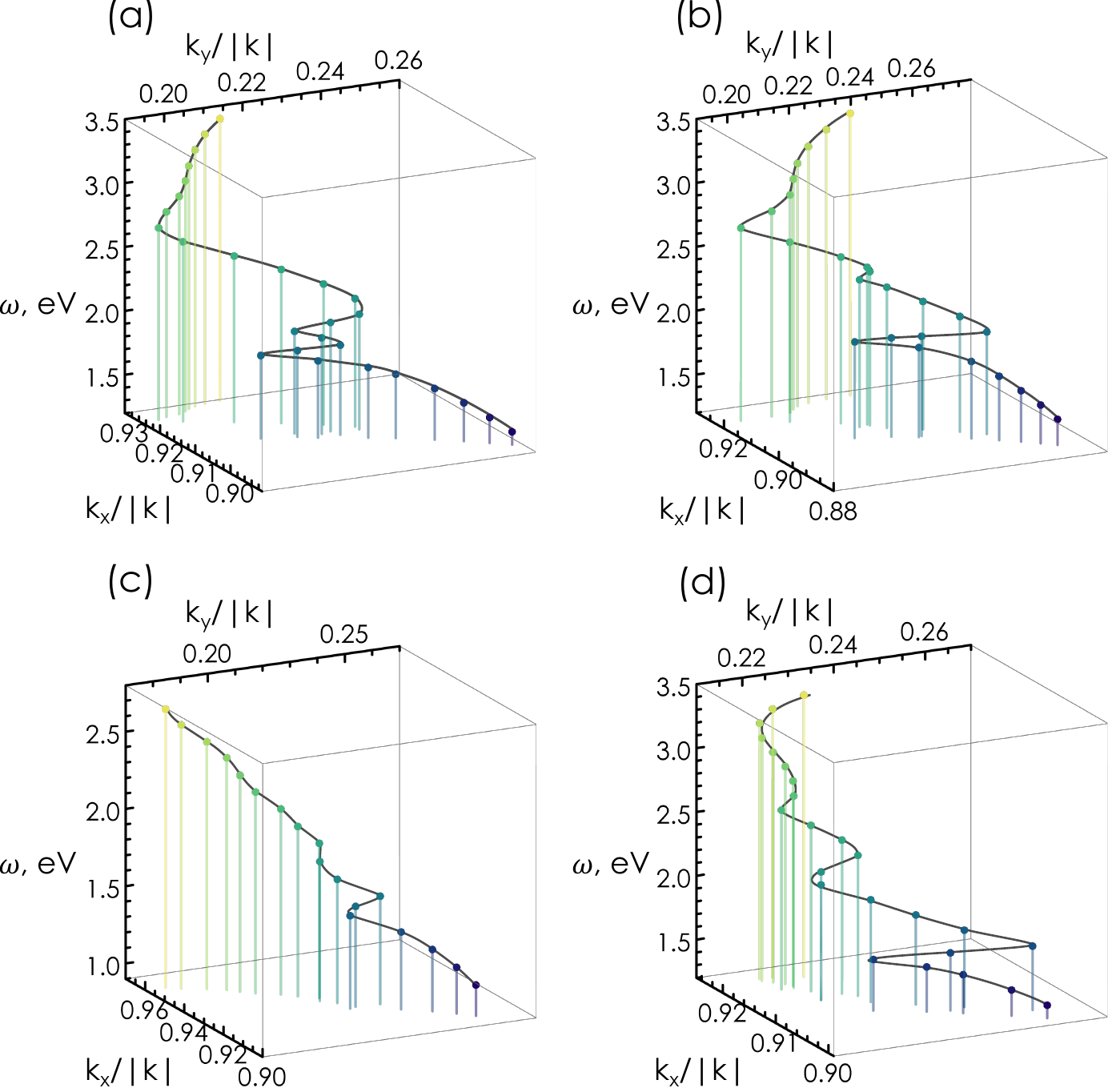}
\caption{
\textbf{Behavior of phase singularities in available uniaxial materials.} 
(a) Trajectory of the $r_{pp}$ reflection phase singularity in the $\bk_\parallel$-$\omega$ space calculated for $\rm{MoS}_2$ at $\alpha=\pi/2$. (b) Same as (a) plotted for $\rm{WS}_2$. (c) Same as (a) plotted for $\rm{MoSe}_2$. (d) Same as (a) plotted for $\rm{WSe}_2$. 
}
\label{fig9}
\end{figure*}

Fig.~\ref{fig8}(b) presents the resulting complex-valued reflection coefficient $r_{pp}$ of a slanted uniaxial material rotated by $\sin{\alpha}=0.6$ around the $y$-axis (other parameters are set to $\eps_\infty = 1$, $f = 1$, $\gamma = 0.1 \omega_0$,  and $\eps_{\parallel} = 2 + 0.5i$). 
Owing to the $xz$ plane of mirror symmetry $r_{pp}(k_x, k_y) = r_{pp}(k_x, -k_y)$.
At the same time, reciprocity ensures $r_{pp}(k_x, k_y) = r_{pp}(-k_x, -k_y)$.
This is equivalent to the simultaneous presence of both $xz$ and $yz$ planes of mirror symmetry in the angular distribution of $r_{pp}$. Thus it suffices to depict the function only in the first quadrant of the parameter space. 
The most important observation is that the violation of axial symmetry by rotating the axis leads to the appearance of an isolated zero in every quadrant of the $k_x$-$k_y$ space, accompanied by phase singularities.
The position of this singularity evolves in the $\bk$-space with frequency, as demonstrated in Fig.~\ref{fig8}(c), which shows the position of the phase singularity at various frequencies at a fixed axis rotation $\alpha$.

As Fig.~\ref{fig8}(b) suggests, the phase gradient distribution around the phase singularity in the slanted uniaxial material (in this particular example) is quite uneven with fast/large gradients in two particular directions.
The model with the resonant Lorentz response \emph{along} the optical axis also features an isolated phase singularity with more even distribution of the phase gradient around the singularity (see Fig. S6 of Supporting Information).

To test the feasibility of this phenomenon with realistic materials, we illustrate the appearance of PSs in the $k_x$-$k_y$ space with a set of four uniaxial transition metal dichalcogenides (TMDCs) WS$_2$, WSe$_2$, MoS$_2$, and MoSe$_2$, whose permittivity tensor is well documented in both the monolayer and bulk form \cite{Ermolaev2021, vyshnevyy2023van}.
We found that reflection PSs in these materials appear only for large rotation angles close to $\pi/2$, which correspond to nearly horizontal optical axis.
Rotating the crystal by $\alpha = \pi/2$ allows us to observe a continuous spectrum of PSs in each of the four TMDCs, Fig. \ref{fig9}.  
The sharp turns in these the trajectories closely correlate with the spectral position of the exciton resonances of these four materials.
It is worth noting that experimental observation of PSs in such systems can be challenging. Manufacturing a sample with a horizontal optical axis and a smooth edge can be a particular challenge, since unwanted scattering from interface irregularities and defects can be an obstacle to observing singularities in the amplitude of the reflected wave.


\section{Conclusion}
To conclude, we have theoretically demonstrated the appearance of phase singularities in light reflection from absorbing uniaxial media. The phase singularities appear as the result of perfect transmission of $p$-polarized light into an absorbing anisotropic medium.
We have analyzed the associated topological charges emerging in the wavelength-incidence angle space, examined the evolution of these singularities with the underlying material parameters, and discussed the relation of this effect to the well-understood phenomenon of Brewster angle in isotropic transparent dielectrics.
Finally, by lowering the symmetry of the problem, we translated this phenomenon into the $k_x$-$k_y$ parameter space, and illustrated the feasibility of this phenomenon with available optically anisotropic materials.
Our results may become valuable for the development of novel optical computing devices relying on topological properties of the optical field, and holography approaches

\section*{Acknowledgments}
Authors acknowledge fruitful discussion with Oleg Kotov, Sergey Dyakov, and thank Georgy Ermolaev for providing  optical constants of anisotropic TMDCs.

\bibliography{Top_Brewster}

\end{document}